\documentstyle[twocolumn,prl,aps,epsf]{revtex}
\begin{document}
\draft
\def \beq{\begin{equation}}
\def \eeq{\end{equation}}
\def \beqarr{\begin{eqnarray}}
\def \eeqarr{\end{eqnarray}}

\twocolumn[\hsize\textwidth\columnwidth\hsize\csname @twocolumnfalse\endcsname

\title{Josephson Effect in Fulde-Ferrell-Larkin-Ovchinnikov 
Superconductors}

\author{Kun Yang and D. F. Agterberg}

\address{
National High Magnetic Field Laboratory and Department of Physics,
Florida State University, Tallahassee, Florida 32310
}

\date{\today}
\maketitle
\begin{abstract}

Due to the difference in the momenta of the 
superconducting order parameters, the Josephson 
current in a Josephson junction between a Fulde-Ferrell-Larkin-Ovchinnikov
(FFLO) superconductor and a conventional BCS superconductor is suppressed.
We show that the Josephson current may be recovered by applying a 
magnetic field in the junction. The field strength and direction 
at which the supercurrent recovery occurs depend
upon the momentum and structure of the order parameter in the FFLO state. 
Thus the Josephson
effect provides an unambiguous way to detect the existence of an
FFLO state, and to measure the momentum of the order parameter.
\end{abstract}
]

It was suggested more than thirty years ago by Fulde and Ferrell\cite{ff},
and Larkin and Ovchinnikov\cite{lo}, that an inhomogeneous superconductor 
with an order parameter that oscillates spatially may be stabilized by a 
large external magnetic or internal exchange field.
Such a Fulde-Ferrell-Larkin-Ovchinnikov (FFLO) state has
never been observed in conventional low-$T_c$ superconductors.
Recently it has attracted renewed interest in the context of organic,
heavy-fermion, and high-$T_c$ cuprate 
superconductors\cite{gloos,yin,norman,rainer,shimahara,murthy,dupuis,modler,tachiki,geg,sh2,maki,samokhin,buzdin,yang,symington,pickett}.
These new classes of superconductors are believed to provide conditions that
are 
favorable to the formation of
FFLO state, because many of them are i) strongly type II superconductors so
that the upper critical field $H_{c2}$ can easily approach the Pauli 
paramagnetic limit; and (ii) layered compounds so that when a magnetic field 
is applied parallel to
the conducting plane, the orbital effect is minimal, and the
Zeeman effect (which is the driving force for the formation of FFLO state)
dominates the physics. Indeed, some experimental indications of the existence
of the FFLO
state have been reported\cite{gloos,modler,geg,symington}.

The main difficulty in the experimental search for the FFLO state is that just
like the BCS state, the FFLO
state is a superconducting state. The distinction between the two is  a
subtle difference in the structure of the superconducting order parameter, 
which is difficult to detect using ordinary experimental techniques.  
Previous experiments have focused on thermodynamic signatures  
of possible phase transitions from the BCS to FFLO state,  
which is believed to be 
first order (see, however, Ref. \onlinecite{rainer}). But such signatures
can be caused by other phase transitions that have nothing to do with the
superconducting order parameter. Thus it is very difficult to establish the
presence of an FFLO state this way without any ambiguity.

In this paper we propose using the
Josephson effect to detect the existence of
FFLO states. 
Our proposal has some similarity in spirit to the
basic ideas behind the
``phase sensitive" experiments\cite{phase} 
that established the predominant $d$-wave
symmetry of the order parameter in cuprate superconductors.
Specifically, we predict: (i) The Josephson current in a Josephson 
junction between a
conventional BCS superconductor and an FFLO superconductor is suppressed, due
to the difference in momenta of the order parameters. (ii) The 
Josephson current may be recovered by applying a properly chosen magnetic field 
in the junction, with field strength and direction depending on the
momentum of the order parameter of the FFLO superconductor; it thus 
provides a way to measure the momentum of the order parameter directly.

In the rest of the paper, we demonstrate the above effects first by using the  
Ginsburg-Landau theory, and then by presenting a microscopic derivation. We  
discuss the experimental implication and feasibility of our proposal 
toward the
end of the paper.

{\em Ginsburg-Landau Theory}.
The effects we predict are most easily demonstrated using a 
Ginsburg-Landau theory. For simplicity we consider a two-dimensional BCS
superconductor, described by a spatially dependent
superconducting order parameter $\Psi_{BCS}({\bf r})$, which is coupled to
a two-dimensional FFLO superconductor, described by an 
order parameter $\Psi_{FFLO}({\bf r})$\cite{note}. 
We consider the two Josephson junction geometries shown in Figure 1.
Since the physics for the two geometries is similar we focus our discussion 
on geometry a) and simply state the results for geometry b). 
For geometry a) the Josephson coupling term in the free
energy takes the form (in the absence of any magnetic field)
\begin{equation}
H_J=-t\int{d^2{\bf r}}[\Psi_{FFLO}^*({\bf r})\Psi_{BCS}({\bf r})+ c.c.].
\label{eq1}
\end{equation}
In the ground state of a BCS 
superconductor, $\Psi_{BCS}({\bf r})= \psi_0$ is a constant. However,
in an FFLO superconductor the order parameter is a superposition
of components carrying finite momenta:
\begin{equation}
\Psi_{FFLO}({\bf r})=\sum_{m}\psi_m e^{i{\bf k}_m\cdot{\bf r}},
\end{equation}
and is oscillatory in space. In the absence of magnetic flux inside the
junction, the total Josephson current is 
\begin{eqnarray}
I_J&=& {\rm Im}\left[t
\int{d^2{\bf r}}\Psi^*_{BCS}({\bf r})\Psi_{FFLO}({\bf r})\right]
\nonumber\\
&=&\sum_m {\rm Im}
\left[t\psi_0^*\psi_m\int{d^2{\bf r}}e^{i{\bf k}_m\cdot{\bf r}}\right].
\end{eqnarray}
Clearly, due to the oscillatory nature of the integrand, the Josephson 
current is suppressed in such a junction. The same result is clearly 
true for geometry b). 

Mathematically, the reason that the Josephson current is suppressed here is
 similar to the suppression of Josephson current by an applied 
magnetic field in an ordinary Josephson junction between two BCS
superconductors\cite{tinkham}. However, the physics is very different:
here the suppression is due to the spatial oscillation of the {\em order
parameter} in the FFLO state, 
while in the case of ordinary Josephson junction in a magnetic
field, the phase of the Josephson tunneling {\em matrix element} is 
oscillatory (in a proper gauge choice). Nevertheless, the mathematical
similarity allows these two effects to
{\em cancel} each other and restore the Josephson current, 
as we demonstrate below.

Consider geometry a).
Imagine applying a parallel
magnetic field ${\bf B}\bot \hat{z}$ parallel to the planes, 
where $\hat{z}$ is a unit vector
along the normal direction of the plane.
Using the gauge ${\bf A}({\bf r})={\bf r}\times {\bf B}$ we have 
${\bf A}({\bf r})=A({\bf r})\hat{z}$. 
The appearance of the ${\bf A}$ field
affects the phase of the Josephson tunneling matrix element only; the in-plane
properties in the two individual superconductors are unaffected because the
${\bf A}$ field is perpendicular to the planes. Specifically,
the Josephson coupling term in the free energy becomes
\begin{equation}
H_J=-t\int{d^2{\bf r}}[e^{{2ieA({\bf r})d\over \hbar c}}
\Psi_{FFLO}^*({\bf r})\Psi_{BCS}({\bf r})+ c.c.],
\end{equation}
and the total Josephson current takes the form
\begin{eqnarray}
I_J&=& {\rm Im}\left[t
\int{d^2{\bf r}}e^{{-2ieA({\bf r})d\over \hbar c}}\Psi^*_{BCS}({\bf r})
\Psi_{FFLO}({\bf r})\right]
\nonumber\\
&=&\sum_m {\rm Im}\left[t\psi_0^*\psi_m\int{d^2{\bf r}}
e^{{-2ieA({\bf r})d\over \hbar c}}e^{i{\bf k}_m\cdot{\bf r}}\right].
\end{eqnarray}
Here $d$ is the distance between the two planes.
It is clear from the equation above that the two oscillatory factors  
cancel each other when
\begin{equation}
{\bf B}={\hbar c\over 2ed}\hat{z}\times {\bf k}_m.
\label{restore}
\end{equation}
At this particular ${\bf B}$, the Josephson current gets partially restored. 
One can determine the momenta of the order parameter in the FFLO state by
searching for the ${\bf B}$'s that restore the Josephson current.

For geometry b) consider applying a perpendicular magnetic field 
${\bf B}\parallel \hat{z}$. Arguments similar to those of 
the last paragraph imply that the Josephson current becomes restored
only when the momentum of the FFLO order parameter is along the
junction (note that typically this momentum will lie along high symmetry
directions of the superconductor). In this case the Josephson current      
is partially restored when
\begin{equation}
{\bf B}=\frac{\hbar c}{2e(d+\lambda_1+\lambda_2)}\hat{z}
\label{eq2}
\end{equation}
where $\lambda_1$ ($\lambda_2$) is the penetration depth of the 
FFLO (BCS) superconductor for the field applied normal 
to the superconducting
plane. The appearance of the factor 
$d+\lambda_1+\lambda_2$ in Eq.~\ref{eq2} is because the magnetic
field is assumed to enter each  superconductor a distance
equal to the penetration depth from the edge of the junction.

{\em Microscopic Derivation}.
These effects can also be demonstrated using a microscopic approach, as we 
illustrate
below. For simplicity and clarity, we take the zero temperature limit, 
and initially consider  
the usual Josephson effect between two BCS superconductors. When two
superconductors are placed in proximity so that electrons can tunnel from one 
to the other, 
the total energy of the system contains a 
term that depends on the phase difference 
$\phi$ between the order parameters of the two superconductors:
\begin{equation}
E_J(\phi)=E_J\cos\phi;
\end{equation}
and the Josephson current is 
\begin{equation}
I_J(\phi)={2eE_J\over\hbar}\sin\phi.
\end{equation}
Thus the calculation of $I_J(\phi)$ is equivalent to the calculation of the
Josephson energy $E_J(\phi)$, which we perform below. We use the 
following tunneling Hamiltonian\cite{josephson,ambegaokar}: 
\begin{equation}
H_T=\sum_{{\bf p}{\bf q}\sigma}(T_{{\bf p},{\bf q}}c_{{\bf p}\sigma}^\dagger
c_{{\bf q}\sigma}+h.c.),
\end{equation}
where ${\bf p}$ and ${\bf q}$ label single electron momentum
eigenstates in the two different superconductors, and 
$\sigma$ is the spin label (this tunneling Hamiltonian
can be used for both geometries since only the matrix elements
$T_{{\bf p},{\bf q}}$ will differ).
The mean field Hamiltonian for a BCS superconductor ($s$-wave)
takes the form:
\begin{eqnarray}
H_l&=&\sum_{{\bf p}\sigma}\epsilon_{\bf p}c_{{\bf p}\sigma}^\dagger
c_{{\bf p}\sigma}+\sum_{\bf p}(\Delta_lc_{{\bf p}\uparrow}^\dagger
c_{-{\bf p}\downarrow}^\dagger+h.c.)\nonumber\\
&=&\sum_{\bf p}E_{\bf p}(\alpha_{\bf p}^\dagger\alpha_{\bf p}
+\beta_{\bf p}^\dagger\beta_{\bf p})+const.,
\label{bcs}
\end{eqnarray}
where $\Delta_l$ is the order parameter of 
the lower superconductor, $\alpha$ and $\beta$ are Bogliubov 
quasiparticle operators, and $E_{\bf p}=\sqrt{\epsilon_{\bf p}^2+|\Delta_l|^2}$
is the quasiparticle energy. When both lower and upper 
superconductors are BCS, a second-order perturbation calculation in 
$H_T$ gives rise to the following expression for the Josephson energy
(we have neglected terms in the second-order perturbation that are independent
of the phase difference $\phi$):
\begin{equation}
E_J(\phi)=-{1\over 2}[\sum_{{\bf p}{\bf q}}{T_{{\bf p},{\bf q}}
T_{-{\bf p},-{\bf q}}\Delta_l^*\Delta_u\over E_{\bf p}E_{\bf q}(E_{\bf p}
+E_{\bf q})}+ c.c.].
\label{ej1}
\end{equation}
In general the tunneling matrix elements $T_{{\bf p},{\bf q}}$ are complex.
A fact that is crucial to the existence of the Josephson effect, as originally
emphasized by Josephson himself\cite{josephson}, is that
in the presence of time-reversal symmetry, $T_{-{\bf p},-{\bf q}}
=T_{{\bf p},{\bf q}}^*$; thus there is no phase oscillation in the summation of
Eq. (\ref{ej1}). Breaking time-reversal symmetry ({\it e.g.} by applying an 
external magnetic field) tends to suppress the 
Josephson effect by introducing oscillatory phases to the terms in Eq.
(\ref{ej1}).

We now turn the discussion to the case where the upper superconductor is in
an FFLO state, while the lower one is still a BCS superconductor. We first
consider zero external magnetic field, and for simplicity
assume the order parameter of the FFLO state has a single momentum 
${\bf k}$. The mean-field Hamiltonian in this case 
takes the form\cite{shimahara,murthy}:
\begin{eqnarray}
&H_u&=\sum_{{\bf q}\sigma}\epsilon_{\bf q}c_{{\bf q}\sigma}^\dagger
c_{{\bf q}\sigma}+E_Z\sum_{{\bf q}}
(c_{{\bf q}\uparrow}^\dagger c_{{\bf q}\uparrow}
-c_{{\bf q}\downarrow}^\dagger c_{{\bf q}\downarrow})\nonumber\\
&+&\sum_{\bf q}(\Delta_uc_{{\bf q}+{\bf k}/2\uparrow}^\dagger
c_{-{\bf q}+{\bf k}/2\downarrow}^\dagger+h.c.)\nonumber\\
&=&\sum_{\bf q}[(E_{\bf q}+Z_{\bf q})\alpha_{\bf q}^\dagger\alpha_{\bf q}
+(E_{\bf q}-Z_{\bf q})\beta_{\bf q}^\dagger\beta_{\bf q}]
+const.,
\label{fflo}
\end{eqnarray}
where $E_Z$ is the Zeeman splitting between the up and down spin electrons, 
and $Z_{\bf q}=E_Z+v_F{\bf q}\cdot{\bf k}/(2|{\bf q}|)$.
The crucial difference between (\ref{bcs}) and (\ref{fflo}) is that in the 
latter case the pairing term creates or annihilates pairs with total momentum
${\bf k}$. A second-order perturbation calculation yields the following 
for the Josephson energy:
\begin{eqnarray}
&E_J(\phi)&=-{1\over 2}[\sum_{{\bf p}{\bf q}}{T_{{\bf p},{\bf q}+{\bf k}/2}
T_{-{\bf p},-{\bf q}+{\bf k}/2}\Delta_l^*
\Delta_u\over E_{\bf p}E_{\bf q}}\nonumber\\
&\times&({\theta(E_{\bf q}-Z_{\bf q})\over E_{\bf p}+E_{\bf q}-Z_{\bf q}}
-{\theta(-E_{\bf q}-Z_{\bf q})\over E_{\bf p}+|E_{\bf q}+Z_{\bf q}|})
+ c.c.].
\label{ej2}
\end{eqnarray}
The expression (\ref{ej2}) looks more complicated than (\ref{ej1}) 
because in the FFLO superconductor  
i) the up and down spin quasiparticles are no longer degenerate and ii) the
ground state is not a vacuum of the quasiparticles. 
The fundamental
difference between (\ref{ej2}) and (\ref{ej1}), however, lies in the fact 
that in (\ref{ej2}) the two tunneling matrix elements forming the product,
$T_{{\bf p},{\bf q}+{\bf k}/2}$ and $T_{-{\bf p},-{\bf q}+{\bf k}/2}$, are
no longer related by time-reversal transformation and therefore are no longer
the complex conjugate of each other. This is due to the fact that the Cooper 
pairs in the FFLO superconductor carry a finite momentum ${\bf k}$. 
As a consequence of this,
the terms in the summation of (\ref{ej2}) have oscillatory phases, and the
Josephson effect is suppressed. 

Consider geometry a) again.
In this case the Josephson energy may be recovered by applying 
a properly chosen
magnetic field parallel to the planes. In the microscopic theory it is 
convenient to choose the following gauge:
\begin{equation}
{\bf A}=-z(\hat{z}\times {\bf B}),
\end{equation}
${\bf A}$ is then parallel to the planes. In this case the gauge 
field does not 
change the tunneling matrix elements. 
Let us assume the lower layer is at $z=0$, while the
upper layer is at $z=d$. Thus ${\bf A}_l=0$ so there is no change
in the BCS superconductor. Also ${\bf A}_u=-d(\hat{z}\times {\bf B})$ is a 
constant gauge field, which shifts the momenta of individual electrons 
in the upper plane by a
constant amount 
\begin{equation}
\delta{\bf k}={ed\over \hbar c}(\hat{z}\times {\bf B}),
\end{equation}
i.e., we need to replace all the ${\bf q}$'s by ${\bf q}+\delta{\bf k}$,
and
$-{\bf q}$'s by $-{\bf q}+\delta{\bf k}$, in
Eqs. (\ref{fflo}) and (\ref{ej2}).
We emphasize that if the upper layer were {\em isolated}, such a shift 
induced by a constant ${\bf A}$ field is a pure gauge transformation that 
has absolutely no physical consequence; the only reason that such a parallel
magnetic field has physical effects is because electrons can tunnel 
from one layer to the other, and thus go through loops that enclose a finite
amount of magnetic {\em flux}.

In particular, when $\delta{\bf k}=-{\bf k}/2$ (as in 
Eq. (\ref{restore})), just as in the BCS superconductor,
the Cooper pairs in the FFLO state carry {\em zero}
momentum (but still finite gauge-invariant center-of-mass {\em velocity}!);
in this case the Josephson energy takes the form:
\begin{eqnarray}
E_J(\phi)&=&-{1\over 2}[\sum_{{\bf p}{\bf q}}{T_{{\bf p},{\bf q}}
T_{-{\bf p},-{\bf q}}\Delta_l^*
\Delta_u\over E_{\bf p}E_{{\bf q}-{{\bf k}\over 2}}}
({\theta(E_{\bf q-{{\bf k}\over 2}}-Z_{{\bf q}-{{\bf k}\over 2}})
\over E_{\bf p}+E_{{\bf q}-{{\bf k}\over 2}}-Z_{{\bf q}-{{\bf k}\over 2}}}
\nonumber\\
&-&{\theta(-E_{{\bf q}-{{\bf k}\over 2}}-Z_{{\bf q}-{\bf k}/2})
\over E_{\bf p}+|E_{{\bf q}-{{\bf k}\over 2}}+Z_{{\bf q}-{{\bf k}\over 2}}|})
+ c.c.].
\end{eqnarray}
Clearly in the above summation we no longer have the oscillatory phases, and
the Josephson energy and current are restored.
Again, similar arguments apply for geometry b) with the conclusions
that the FFLO momentum must lie along the junction to restore
the Josephson effect and, if this is the case, the restoration will
occur at the field given by Eq.~\ref{eq2}.  

{\em Discussion.} While arriving at the same basic conclusions, 
the microscopic and macroscopic derivations do not 
lead to results that are equivalent in all details. 
This is due to the implicit assumption 
made in the macroscopic derivation that the Josephson coupling is uniform 
throughout the junction. This assumption was not made at the microscopic
level; if made, it will restrict the form of the tunneling 
matrix elements $T_{\bf p,q}$.   
To illustrate this consider an infinite junction. In this case 
translational invariance in the junction requires $T_{\bf p,q}=e^{i({\bf p}-
{\bf q})\cdot {\bf t}_j} T_{\bf {p,q}}$ where ${\bf t}_j$ is an arbitrary
translation
vector in the junction. Implementing this constraint in Eq.~(\ref{ej1}) gives 
$E_J(\phi)\propto  
\delta_{{\bf k},0}$ which is also the result of Eq.~(\ref{eq1}) 
after the spatial  
integration is performed. 

In our discussion we have assumed that the
superconductors are in the clean limit, so that we can label single electron
states by their momenta, and the order parameter of the FFLO superconductor
has either a single momentum or is a linear superposition of components with a 
discrete set of momenta. The presence of disorder will change the situation.
In this case it is still true that the Josephson effect between a BCS 
superconductor and an FFLO superconductor is suppressed, as in the latter
Cooper pairs are {\em not} formed between single electrons states that are
related by time-reversal transformation. However, the recovery  of the
Josephson effect by applying a magnetic field will be decreased, 
since in this case the
order parameter in the FFLO state is formed by superpositions of components
with momentum that varies continuously; while a magnetic field can
only
recover the contribution to the Josephson energy of a single component with a
fixed momentum. Thus experimental studies of the effects we predict should be
performed in clean superconductors. 

Ideally, we would like to apply this theory to FFLO superconductors that
are stabilized by an internal exchange field, rather than an external magnetic
field. This is because in the latter case, unless the system is truly
two-dimensional, the orbital effect of the
magnetic field that penetrates into the superconductor
will induce vortices, which will complicate the analysis.
We note that while experimentally
the 
potential candidates for FFLO superconductors reported so far
are stabilized by an external 
field\cite{gloos,modler,geg,symington},
recent theoretical studies suggest that an FFLO state may form in 
the ferromagnetic superconductor
RuSr$_2$GdCu$_2$O$_8$\cite{pickett} without the benefit of an
external magnetic field.
Despite the disclaimer made above, we expect that the above
results will be robust in the presence of a weak orbital effect,
for example in 
quasi 2D systems where there is very weak 
interlayer coupling. In this case the orbital effects will be minimal
when the magnetic field is applied in the layers.  

In conclusion we have shown that while the Josephson current between
a BCS and an FFLO superconductor is suppressed, it is restored by
applying a magnetic field in the junction. The field strength and
direction at which the Josephson current recovers depend upon the
momenta of the FFLO order parameter, allowing the momenta to be measured.
The observation of a field restored Josephson effect will provide 
unambiguous evidence for an FFLO phase.

This work was supported by NSF DMR-9971541 and the A. P. Sloan Foundation
(KY), and by NSF DMR-9527035 and the State of Florida (DFA).

\begin{figure}
\epsfxsize=2.5 in
\epsfbox{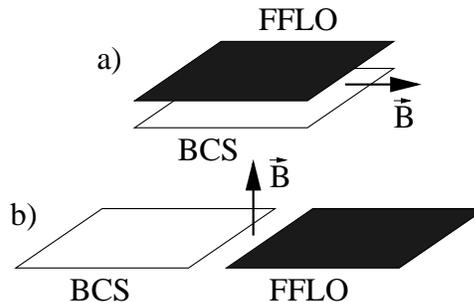}
\caption{Josephson junction geometries considered in this article.
For geometry a) the magnetic field in the junction 
is applied in the plane of the
superconductors; while for geometry b) the magnetic field is applied
perpendicular to the plane of the superconductors, but confined inside the
junction area.} 
\label{fig1}
\end{figure}
\end{document}